\documentclass[%
reprint,
superscriptaddress,
amsmath,amssymb,
prl,
]{revtex4-2}
\usepackage{color}
\usepackage{float}
\usepackage[pdftex]{graphicx}
\usepackage{dcolumn}
\usepackage{bm}
\usepackage{siunitx}
\usepackage{comment}
\usepackage{soul}
\usepackage{braket}
\usepackage{dcolumn}

\usepackage{hyperref}
\hypersetup{
    unicode=false,          
    pdftoolbar=true,        
    pdfmenubar=true,        
    pdffitwindow=false,     
    pdftitle={Broadband parametric amplification for multiplexed SiMOS quantum dot signals},    
    pdfauthor={Victor Elhomsy},     
    pdfsubject={},   
    pdfcreator={Victor Elhomsy},   
    pdfproducer={},  
    pdfkeywords={,} {} {}, 
    pdfnewwindow=true,      
    colorlinks=false,       
    linkcolor=red,          
    citecolor=green,        
    filecolor=magenta,      
    urlcolor=cyan           
}

\usepackage{comment}

\DeclareUnicodeCharacter{2009}{\,} 
\DeclareSIUnit{\belmilliwatt}{Bm}
\DeclareSIUnit{\dBm}{\deci\belmilliwatt}

\begin{document}
\bibliographystyle{apsrev4-1}

\preprint{APS}
\title{Broadband parametric amplification for multiplexed SiMOS quantum dot signals}

\author{Victor Elhomsy}
\email{victor.el-homsy@neel.cnrs.fr}
\affiliation{Universit\'e Grenoble Alpes, CNRS, Grenoble INP, Institut N\'eel, 38042 Grenoble, France}

\author{Luca Planat}
\affiliation{Universit\'e Grenoble Alpes, CNRS, Grenoble INP, Institut N\'eel, 38042 Grenoble, France}

\author{David J. Niegemann}
\affiliation{Universit\'e Grenoble Alpes, CNRS, Grenoble INP, Institut N\'eel, 38042 Grenoble, France}

\author{Bruna Cardoso-Paz}
\affiliation{Universit\'e Grenoble Alpes, CNRS, Grenoble INP, Institut N\'eel, 38042 Grenoble, France}

\author{Ali Badreldin}
\affiliation{Universit\'e Grenoble Alpes, CNRS, Grenoble INP, Institut N\'eel, 38042 Grenoble, France}

\author{Bernhard Klemt}
\affiliation{Universit\'e Grenoble Alpes, CNRS, Grenoble INP, Institut N\'eel, 38042 Grenoble, France}

\author{Vivien Thiney}
\affiliation{Universit\'e Grenoble Alpes, CNRS, Grenoble INP, Institut N\'eel, 38042 Grenoble, France}

\author{Renan Lethiecq}
\affiliation{Universit\'e Grenoble Alpes, CNRS, Grenoble INP, Institut N\'eel, 38042 Grenoble, France}

\author{Eric Eyraud}
\affiliation{Universit\'e Grenoble Alpes, CNRS, Grenoble INP, Institut N\'eel, 38042 Grenoble, France}

\author{Matthieu C. Dartiailh}
\affiliation{Universit\'e Grenoble Alpes, CNRS, Grenoble INP, Institut N\'eel, 38042 Grenoble, France}

\author{Benoit Bertrand}
\affiliation{Universit\'e Grenoble Alpes, CEA, Leti, F-38000 Grenoble, France}

\author{Heimanu Niebojewski}
\affiliation{Universit\'e Grenoble Alpes, CEA, Leti, F-38000 Grenoble, France}

\author{Christopher B{\"a}uerle}
\affiliation{Universit\'e Grenoble Alpes, CNRS, Grenoble INP, Institut N\'eel, 38042 Grenoble, France}

\author{Maud Vinet}
\affiliation{Universit\'e Grenoble Alpes, CEA, Leti, F-38000 Grenoble, France}

\author{Tristan Meunier}
\affiliation{Universit\'e Grenoble Alpes, CNRS, Grenoble INP, Institut N\'eel, 38042 Grenoble, France}

\author{Nicolas Roch}
\affiliation{Universit\'e Grenoble Alpes, CNRS, Grenoble INP, Institut N\'eel, 38042 Grenoble, France}

\author{Matias Urdampilleta	}
\email{matias.urdampilleta@neel.cnrs.fr}
\affiliation{Universit\'e Grenoble Alpes, CNRS, Grenoble INP, Institut N\'eel, 38042 Grenoble, France}

\date{\today}

\begin{abstract}
Spins in semiconductor quantum dots hold great promise as building blocks of quantum processors. Trapping them in SiMOS transistor-like devices eases future industrial scale fabrication. Among the potentially scalable readout solutions, gate-based dispersive radiofrequency reflectometry only requires the already existing transistor gates to readout a quantum dot state, relieving the need for additional elements. In this effort towards scalability, traveling-wave superconducting parametric amplifiers significantly enhance the readout signal-to-noise ratio (SNR) by reducing the noise below typical cryogenic low-noise amplifiers, while offering a broad amplification band, essential to multiplex the readout of multiple resonators. In this work, we demonstrate a $\SI{3}{\giga\hertz}$ gate-based reflectometry readout of electron charge states trapped in quantum dots formed in SiMOS multi-gate devices, with SNR enhanced thanks to a Josephson traveling-wave parametric amplifier (JTWPA). The broad, tunable $\SI{2}{\giga\hertz}$ amplification bandwidth combined with more than $\SI{10}{\decibel}$ ON/OFF SNR improvement of the JTWPA enables frequency and time division multiplexed readout of interdot transitions, and noise performances near the quantum limit. In addition, owing to a design without superconducting loops and with a metallic ground plane, the JTWPA is flux insensitive and shows stable performances up to a magnetic field of $\SI{1.2}{\tesla}$ at the quantum dot device, compatible with standard SiMOS spin qubit experiments.
\end{abstract}

\maketitle

\section{I. Introduction}

Since seminal proposals \cite{LossDivin, Kane1998}, spins trapped in semiconductor quantum dots have attracted a lot of attention in the prospect to be used as bits of information for quantum computation. The SiMOS industrial platform could implement this technology, with the hope to leverage the well-established microelectronics foundry in order to scale up from a few lab qubits to a processor containing billions of them \cite{Maurand2016, GZ2021}.

The readout part of a quantum algorithm must meet this scalability criterion. In semiconductor quantum dots spin qubit experiments, spin state readout relies on spin-to-charge conversion techniques, where the charge susceptibility depends on the spin state of the qubit \cite{elzerman2004, ono2002}. Such techniques can be realized by embedding the device in a radiofrequency (RF) resonator, and probing the shift in resonance frequency due to a change in charge susceptibility of the quantum dot \cite{schoelkopf1998, Vigneau2023}. A scalable example of such techniques relies on using resonators connected to already-existing metallic gates and measuring a capacitance signal, when a non-zero charge susceptibility arises at a charge transition only if tunneling is allowed \cite{petersson2010, colless2013, betz2015, house2015, urdampilleta2015, GZ2015, GZ2016, vanderheijden2018, ahmed2018, crippa2019, zheng2019, west2019, lundberg2020, hogg2023}. This method requires small RF power to avoid populating excited states, averaging out the signal with modulation, and heating. This in-situ readout relieves the need for nearby reservoirs or additional charge sensors.

In such cryogenic semiconductor experiments, the amplitude of the signal of interest is so small that multiple amplification stages are needed between the output of the device and room temperature readout electronics. Superconducting parametric amplifiers \cite{aumentado2020}, that can be placed at the coldest stage of a cryostat, offer a typical $10-\SI{20}{\decibel}$ gain and a noise temperature an order of magnitude lower than typical semiconductor low noise amplifiers operating at the \SI{4}{\kelvin} stage \cite{stehlik2015, schaal2020, schupp2020, dejong2021, oakes2023}. Various approaches are being developed, either based on arrays of Josephson junctions or SQUIDs \cite{castellanos2008, vijay2009, bergeal2010, macklin2015, planat2020, roudsari2023}, high kinetic inductance superconductors \cite{chaudhuri2017, goldstein2020, malnou2021, parker2022, vine2022, xu2023, khalifa2023}, semiconductors \cite{kass2023}, hybrid devices \cite{phan2022} or graphene \cite{butseraen2022, sarkar2022}, with a trade-off between low bandwidth due to resonant processes, little material resilience to external magnetic fields or to incoming power, or an added noise further from the quantum limit.

Among these solutions, traveling-wave parametric amplifiers \cite{esposito2021} seem more suitable to scale up qubit readout beyond the bandwidth constraints of resonant architectures, at the cost of a more complex nanofabrication and challenges in impedance and phase matching. Indeed, one way to tackle the scalability challenge is to have a single readout line, and to perform parallel readout thanks to multiplexing \cite{jerger2012, jeffrey2014, dejong2021, Ruffino2022}. Simultaneous multiplexed readout of semiconductor spin qubits requires large bandwidth, high compression point and magnetic field resilience, while keeping the noise level as low as possible. Besides, in this scalability context, typical circuit quantum electrodynamics resonators have footprints comparable to the signal wavelength, in the few $\SI{}{\centi\meter}$ range for $\SI{}{\giga\hertz}$ experiments, limiting their on-chip integration when going to more and more qubits.

In this article, we use a Josephson-junction based traveling-wave superconducting parametric amplifier (JTWPA) to probe the impedance signal of quantum dots embedded in small-footprint lumped-element resonators. We first characterize the JTWPA in the frequency regime corresponding to our experiments: we find nominal gain, bandwidth and compression point compared to its usual regime of operation at higher frequencies, and a resilience to the externally applied magnetic field at the position of the quantum dot device. We then characterize the noise performance of the readout, and find a JTWPA added noise close to the quantum limit of an amplification process. In a third part, we demonstrate its use in dispersive charge susceptibility measurements at interdot charge transitions in SiMOS double quantum dots. We finally leverage its broadband gain to multiplex the readout of two interdot signals.

\section{II. Setup and device description}

\begin{figure}%
\includegraphics[width=\columnwidth]{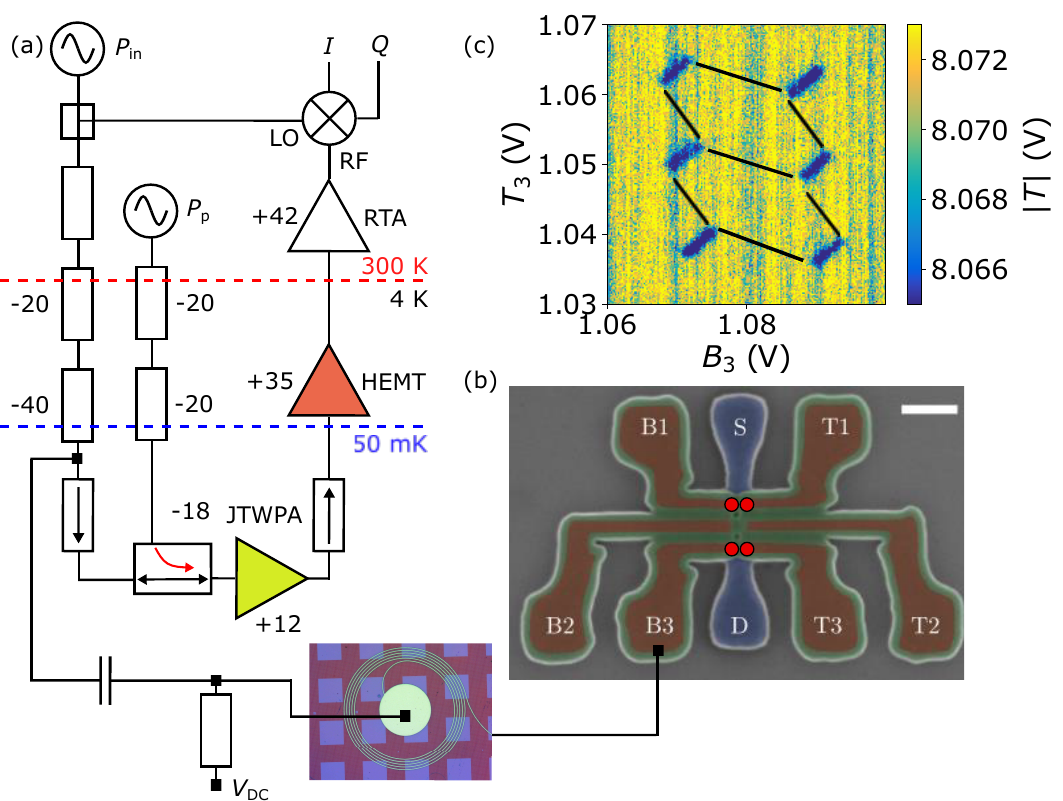}
\caption{(a) Electronic and cryogenic setup. An RF tone $P_\text{in}$ is split in two at room temperature (RT). One part serves as local oscillator (LO) for demodulation, the other goes through multiple attenuation stages between RT and the coldest stage in the cryostat ($\SI{50}{\milli\kelvin}$). A second RF source provides energy to the JTWPA through the pump tone $P_\text{p}$, which is fed to the JTWPA via a directional coupler. $P_\text{in}$ goes through a PCB feedline and an isolator to protect the device from pump tone reflection. The resulting RF signal is amplified by the JTWPA, then goes through a second isolator that prevents reflected $\SI{4}{\kelvin}$ radiation from polluting the signal. It is then amplified by the HEMT amplifier anchored to the $\SI{4}{\kelvin}$ stage and by room temperature amplifiers (RTA). It is finally demodulated, and both $I$ and $Q$ quadrature are further amplified, filtered and read out in a homemade ADC (not shown in the picture). (b) The three split-gate device studied in this work. Three pairs of metal gates (brown) are deposited on top of a silicon nanowire (blue), with doped electron reservoirs source (S) and drain (D). Red points are the position of the quantum dots of interest, below gates $\text{T}_1$, $\text{B}_1$, $\text{T}_3$ and $\text{B}_3$. In the picture, gate $\text{B}_3$ is connected to an on-chip spiral inductor to form the $LC_\text{p}$ resonator with the device and PCB parasitic capcitance $C_\text{p}$. The RF tone and low frequency control signal $V_\text{DC}$ get to $\text{B}_3$ through a bias tee on the PCB. (c) Typical stability diagram of a double quantum dot (DQD). Gates $\text{T}_3$ and $\text{B}_3$ accumulate electrons in the Si channel. The signal mapped here is the modulus of RF transmission $T$ through the setup. Yellow areas are zones of charge stability. Blue lines highlight interdot charge transitions, where electrons can hop between dots below $\text{T}_3$ and $\text{B}_3$.}
\label{fig:setup}
\end{figure}

The signal of interest in this work is the dispersive change in resonance frequency of a lumped-element resonator in which our device is embedded, at the coldest stage ($\SI{50}{\milli\kelvin}$) of a homemade dilution refrigerator, see Fig.\ref{fig:setup}(a). We send a radiofrequency (RF) tone of power $P_\text{in}$ through successive stages of attenuation. On the cold plate, the RF tone goes through a feedline on the PCB where the device is glued. The device is connected to the feedline in a "hanger" geometry, through a bias tee combining low frequency control electronics (gate voltages), and the RF tone of interest. After the feedline, the output signal goes through an isolator to protect the device from reflections of the JTWPA pump tone $P_{\text{p}}$. It then goes through the JTWPA, a second isolator to protect the JTWPA from $\SI{4}{\kelvin}$ radiation, a high-electron-mobility transistor (HEMT) amplifier anchored at the $\SI{4}{\kelvin}$ stage, further room temperature amplifiers, and it is finally demodulated, where we get both $I$ and $Q$ quadratures.

To form the resonator, we integrate niobium superconducting spiral inductors directly on-chip \cite{Niegemann2022}, in a post-CMOS fabrication process \cite{Klemt2023, Yu2023}. This forms a $LC_\text{p}$-resonator due to the parasitic capacitance $C_\text{p}$ from the printed circuit board, the bonding pads and the device structure. Patterning the inductor directly on-chip enabled to reduce $C_\text{p}$ compared to inductors fabricated on a separate chip and wire bonded to the device. This results in a resonance frequency closer to the range where superconducting amplifiers operate. The principle of this readout method is to send $P_\text{in}$ on a resonance flank, so that a small shift in resonance frequency will result in a sensitive variation of transmitted signal through the feedline.

The devices studied in this work are based on $\SI{80}{\nano\meter}$-wide silicon nanowire transistors from $\SI{28}{\nano\meter}$ fully depleted silicon-on-insulator (FD-SOI) technology \cite{silvano2016}. On top of a $\SI{10}{\nano\meter}$ thick and $\SI{80}{\nano\meter}$ wide nanowire, we deposit $\SI{6}{\nano\meter}$ SiO$_2$ gate dielectric, $\SI{5}{\nano\meter}$ TiN and $\SI{50}{\nano\meter}$ poly-Si metal for the gate layer. In the device depicted in Fig.\ref{fig:setup}(b), three pairs of split gates are deposited with $\SI{40}{\nano\meter}$ gate width, $\SI{40}{\nano\meter}$ gate pitch, and $\SI{40}{\nano\meter}$ split gate separation. The gate pitches are filled with SiN spacers. The device is then buried in $\SI{200}{\nano\meter}$ of SiO$_2$, with W vias to connect the gates and the reservoirs of the transistor.

At cryogenic temperatures, the gates of the device can be used to trap individual charges in the active channel of the transistor. With more and more positive gate voltage, electrons accumulate in "corner dots" \cite{silvano2016} due to the wrap-around shape of the gate.  We can probe the population of a double quantum dot (DQD) below a couple of split gates by performing typical stability diagrams like in Fig.\ref{fig:setup}(c). When an electron can tunnel between the dots below gates $\text{T}_3$ and $\text{B}_3$, the charge of the system is degenerate, which creates a capacitive shift of resonance frequency. The transmitted $(I,Q)$ signal on these "interdot" charge transition lines is different from regions of stable charge occupancy. For the device shown here, we linked one resonator to each pair of split gates. More precisely, an inductor is connected to $\text{T}_1$, to $\text{T}_2$ and to $\text{B}_3$ (only the one linked to $\text{B}_3$ is shown on the figure).

The main figure of merit of such a readout protocol is the signal-to-noise ratio, which quantifies the contrast between interdot zones and stable zones on these diagrams. It can be written:
\begin{equation}
    \text{SNR} = |\Delta T|^2 \frac{P_{\text{in}}}{P_\text{n}}
\label{eq:snr}
\end{equation}
where $|\Delta T|^2$ is the shift in RF transmission due charge degeneracy ($T=\sqrt{I^2+Q^2}$), $P_\text{in}$ is the input RF power, and $P_\text{n}$ is the noise power. $|\Delta T|^2$ can be enhanced thanks to a better resonator quality factor or to a better coupling between gates and quantum dots \cite{vonhorstig2023}. In our context, $P_\text{in}$ is limited by interdot power broadening, when the RF modulation becomes larger than the interdot linewidth and averages it out. In practice, tunnel couplings in the range of few tens of $\SI{}{\micro\electronvolt}$ limit $P_\text{in}$ to about $\SI{-100}{\deci\belmilliwatt}$, and depend on the resonator quality factor and on the gate lever-arms.

$P_\text{n}$ contains the information on the noise that hinders signal readout. In the case of an amplification chain, it is dominated by the added noise of the first amplifier \cite{pozar2011}, and the subsequent sources of noise are divided by the gain of this first amplifier. This is the cornerstone of readout performance, and in our case this amplifier will be the JTWPA.

\begin{figure}%
\includegraphics[width=0.75\columnwidth]{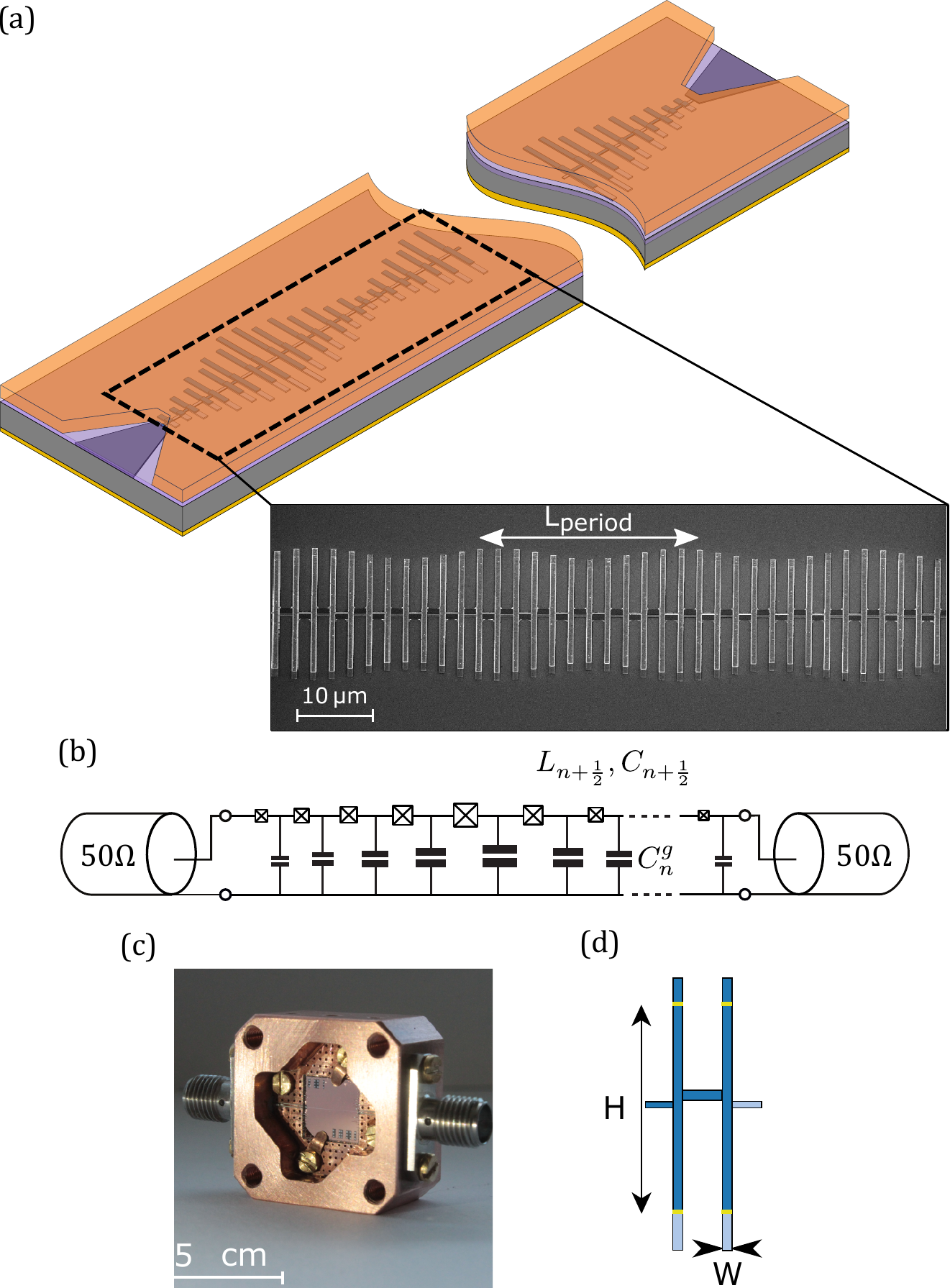}
\caption{The Josephson traveling-wave parametric amplifier (JTWPA), adapted from \cite{planat2020}. (a) Schematic of the JTWPA. An array of aluminum Josephson junctions is fabricated on top of a silicon substrate. They are then covered by a thin ALD-grown alumina layer (purple), and a thick copper layer (orange) used as a top ground to enforce impedance matching. Below, we show a SEM image of three periods of a spatially modulated array of junctions. (b) Electrical sketch of the JTWPA. Crosses in squares represent Josephson junctions, consisting in a capacitor and a non-linear inductor in parallel. (c) Picture of the sample. The chip is the square in the middle, anchored to a PCB in a copper case. (d) Schematic of one Josephson junction. Here, only the junctions height $H$ (the overlap between top and bottom aluminum layers) is modulated in space, not their width $W$.
}
\label{fig:twpa}
\end{figure}

The JTWPA used here consists in a chain of single aluminum Josephson junctions, whose characteristic heights are periodically modulated, (see Fig.\ref{fig:twpa}(a-b)). It is a single junction adaptation of the SQUID-based amplifier described in references \cite{planat2019, planat2020} that give details on the fabrication process, and a quantitative description of the amplification phenomenon. The chip containing the JTWPA is depicted at the center of Fig.\ref{fig:twpa}(c), plugged to a copper printed circuit board, and mounted into a copper box. Thermal anchoring is ensured by fixing the JTWPA to a copper cold-finger screwed on the cold plate. This cold-finger is about $\SI{10}{\centi\meter}$ long, to place the JTWPA as far as possible from the center of the coil used to generate the magnetic field meant to lift the degeneracy of spin states of electrons trapped in the transistor device. Except some turns of aluminum tape around the copper box, we do not use any other protection against magnetic field.

\section{III. JTWPA performance in our regime}

Superconducting parametric amplifiers are usually designed to be operated in the $4-\SI{12}{\giga\hertz}$ range \cite{aumentado2020, esposito2021}. In contrast, usual semiconductor spin qubit experiments using RF relectometry happen below $\SI{1}{\giga\hertz}$ because of the parasitic capacitance arising from bonding pads when using surface mounted inductors or inductors patterned on a separate chip: the spiral inductor gives a few tens of $\SI{}{\nano\henry}$, and the parasitic capacitance lies between $0.5-\SI{1}{\pico\farad}$. Patterning the inductor on chip as described in the last section enabled to go higher in frequency and to reach the $2-\SI{4}{\giga\hertz}$ range.

\begin{figure}%
\includegraphics[width=\columnwidth]{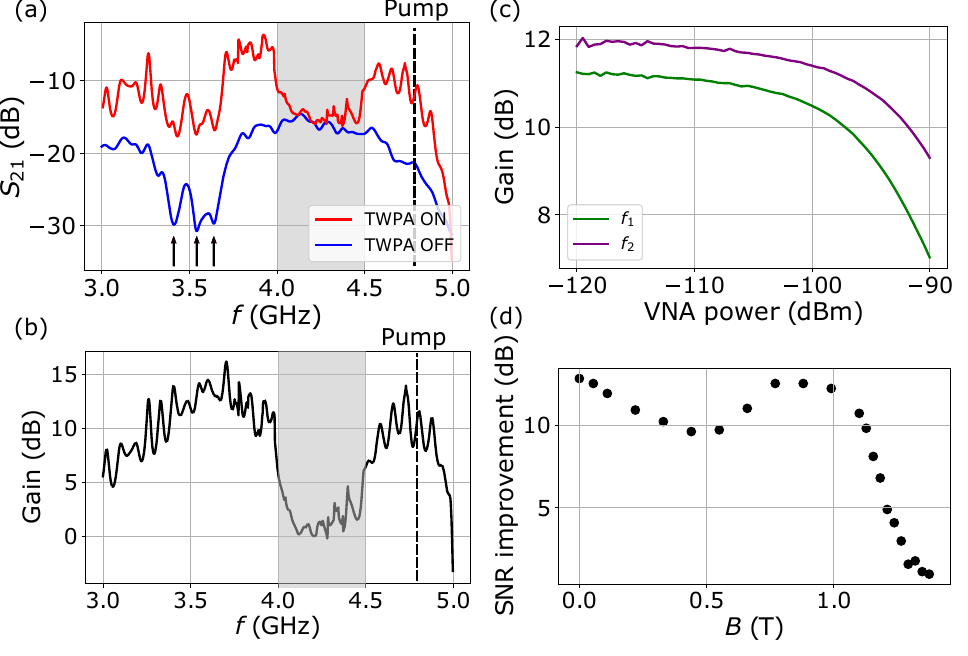}
\caption{Characteristics of the JTWPA. (a) Transmission through the whole RF setup. The blue curve, with JTWPA OFF (i.e pump RF source output OFF), shows the three resonances of the resonators linked to $\text{T}_1$, $\text{T}_2$ and $\text{B}_3$ respectively around $\SI{3.42}{\giga\hertz}$, $\SI{3.51}{\giga\hertz}$ and $\SI{3.52}{\giga\hertz}$, indicated by arrows. The red curve, with JTWPA ON, adds the JTWPA gain to the blue curve. We pinpoint the pump tone at $\SI{4.76}{\giga\hertz}$. (b) JTWPA gain as a function of incoming RF frequency (difference between JTWPA ON and OFF curves of panel (a)). (c) Compression point: JTWPA gain as a function of RF power at the input of the JTWPA, taken at frequencies $f_{1,3}$ corresponding to optimal readout frequencies for multiplexed data in last section. (d) JTWPA SNR improvement as a function of the externally applied magnetic field at the position of the quantum dot device. We see a resilience of the JTWPA noise performance up to $\SI{1.2}{\tesla}$ felt by the device.
}
\label{fig:JTWPA_charac}
\end{figure}

Fig.\ref{fig:JTWPA_charac}(a) shows the transmission $S_{21}$ of the setup, comparing JTWPA ON and OFF. The setup is the same as described before, except that a vector network analyzer (VNA) provides $P_\text{in}$ and retrieves the signal after the last amplifier. The JTWPA OFF transmission shows the dips of the three resonators, plugged to $\text{T}_1$, $\text{T}_2$ and $\text{B}_3$, around $\SI{3.41}{\giga\hertz}$, $\SI{3.65}{\giga\hertz}$ and $\SI{3.53}{\giga\hertz}$ respectively. The JTWPA ON $S_{21}$ shows different zones, which are better understood by looking directly at the gain (difference between ON and OFF curves), see Fig.\ref{fig:JTWPA_charac}(b). The spatial periodic modulation of the JTWPA opens a gap in its dispersion relation, like in a photonic crystal \cite{planat2020}. It appears here after $\SI{5}{\giga\hertz}$. The transmission gap around $\SI{4.3}{\giga\hertz}$ is the symmetric of this photonic gap compared to the pump frequency, sent at $\SI{4.73}{\giga\hertz}$ (RF source for the pump tuned at $-\SI{5.5}{\deci\belmilliwatt}$). Below $\SI{4}{\giga\hertz}$ is the area of interest for us: the JTWPA gives between $12-\SI{15}{\decibel}$ bare ON/OFF gain over $\SI{0.5}{\giga\hertz}$, and tunable over a broad $\SI{2}{\giga\hertz}$ band up to the gap, by tuning the pump frequency and power. This contrasts with resonant Josephson amplifiers that need to be finely tuned near resonance. Wiggles in the gain profile come from impedance mismatch between the JTWPA and RF circuitry connected to it.

Fig.\ref{fig:JTWPA_charac}(c) shows JTWPA gain as a function of RF power at the JTWPA input, for frequencies $f_1 = \SI{3.465}{\giga\hertz}$ and $f_3 = \SI{3.525}{\giga\hertz}$, in the same pump configuration than for Fig\ref{fig:JTWPA_charac}(a,b). These correspond to optimal reflectometry working points to readout interdot charge transitions in ($\text{T}_1/\text{B}_1$) and ($\text{T}_3/\text{B}_3$) DQDs. Both curves lose $\SI{1}{\decibel}$ at around $\SI{-100}{\deci\belmilliwatt}$. This gain compression is 2-3 orders of magnitude higher than typical resonant Josephson parametric amplifiers and enables multiplexed readout of these two interdot transitions, as will be demonstrated in the last section of this article.

We finally turn to the magnetic field resilience of the JTWPA. In the context of semiconductor spin qubits, quantum dots in these SiMOS devices are meant to trap individual electrons, in order to coherently manipulate their spins. In such experiments, a magnetic field is needed to lift the degeneracy of spin states \cite{elzerman2004, Niegemann2022}. Fig.\ref{fig:JTWPA_charac}(d) shows the SNR improvement of the JTWPA with respect to an externally applied magnetic field at the position of the SiMOS device. The reported magnetic field is the one felt by the device at the center of an external coil, not the one felt by the JTWPA, which is likely to be two orders of magnitude smaller given its position and orientation in the cryostat. We see that for fields up to $\SI{1.2}{\tesla}$, the SNR improvement stays in a $[\SI{10}{\decibel};\SI{13}{\decibel}]$ window, and begins to drop for higher fields. This resilience of the JTWPA to the magnetic field is explained by the single-junction nature of the JTWPA, compared to SQUID-based amplifiers that are highly sensitive to the magnetic flux, and also by the non-superconducting metallic top ground. This mu-metal free, easy to implement magnetic field resilience encourages the use of JTWPAs in usual semiconductor spin qubit experiments.

\section{IV. System noise and JTWPA added noise}

\begin{figure}%
\includegraphics[width=\columnwidth]{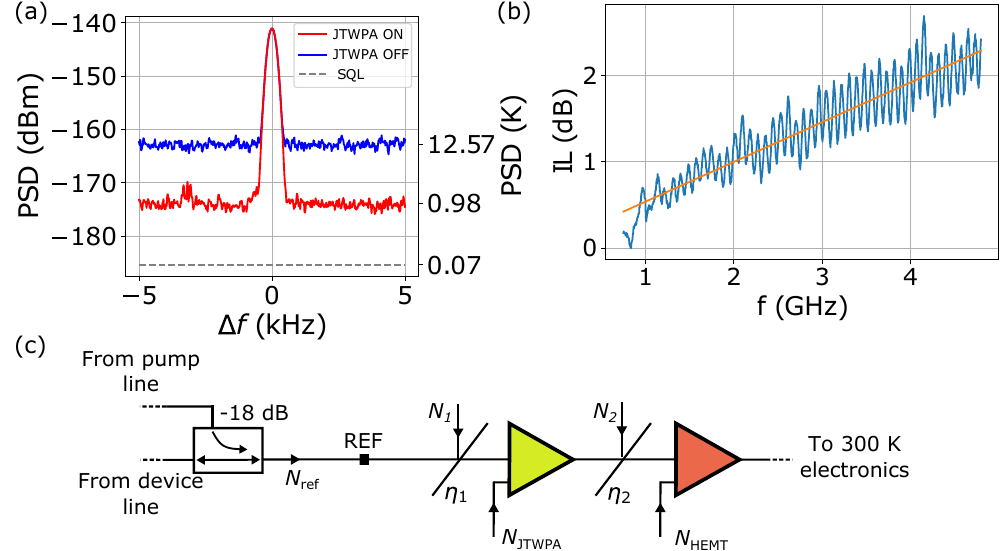}
\caption{(a) Spectrum analyzer measurement of a tone sent at $f=\SI{2.911}{\giga\hertz}$, JTWPA ON (red) and OFF (blue), offsetted by the amplification chain. The gray dashed line shows the standard quantum limit of added noise for an amplification process, for comparison with the noise floors. (b) JTWPA insertion loss. The blue curve shows measured data, the orange line is a fit to $IL = 0.46\times f[\text{GHz}] + 0.08$. (c) Model for noise estimation, adapted from \cite{blais2021}. $N_{\text{JTWPA}}$ and $N_{\text{HEMT}}$ are the added noise of the JTWPA and the HEMT. The RF circuitry before the JTWPA, and between the JTWPA and the HEMT, is modelled as lossy beam splitters of transmissivities $\eta_{1,2}$. $N_{1,2}$ are vacuum noise incoming at the beam splitters. "REF" is the reference point where we express the noise temperature: the output gain $G_{\text{out}}$ in eq.\ref{eq:Pn} is the sum of all gains and attenuations beyond this point.
}
\label{fig:noise}
\end{figure}

To assess the noise performance of the JTWPA-enhanced measurement, we perform the following measurement on a similar device, with a resonance frequency at $f=\SI{2.911}{\giga\hertz}$ (pump frequency $\SI{4.269}{\giga\hertz}$ and RF source power $\SI{-7.26}{\deci\belmilliwatt}$). We provide $P_\text{in}$ with an external RF source tuned at $f$, and we measure the tone coming out of the cryostat with a spectrum analyzer (SA), with a $B=\SI{300}{\hertz}$ resolution bandwidth. In Appendix 1, we plot the resulting signal on a $\SI{10}{\kilo\hertz}$ window, JTWPA ON and OFF. The usual way to represent these data is to offset them by the total amplification of the RF output line, like in Fig.\ref{fig:noise}(a). Appendix 1 details the estimation of the output line. On such a plot, we can directly read a +$\SI{11.1}{\decibel}$ SNR improvement as the difference between the noise floors ON and OFF. This was obtained on a different device than before, and with a different pump configuration. It demonstrates the JTWPA tunability to adapt to different optimal readout frequencies.

An important point to stress here is that in our setup, the "JTWPA OFF" curves are obtained simply by not pumping the JTWPA while measuring. But we don't remove it from the chain. This implies that the data with JTWPA OFF are artificially degraded due to the JTWPA insertion loss. To account for this, we estimated the insertion loss in a different setup, in the frequency window between the low-frequency cutoff of the isolators and the JTWPA photonic gap. We measured the full RF transmission through the cryostat, first with a $\SI{50}{\ohm}$ short on the cold plate, then by replacing the short by the unpumped JTWPA. We plot the difference between these measurements on Fig.\ref{fig:noise}(b). A linear fit gives a slope of $\SI{0.46}{\decibel\per\giga\hertz}$. At the frequency where Fig.\ref{fig:noise}(a) was taken, the insertion loss is $\SI{1.26}{\decibel}$. We took it into account in the SA plot, by subtracting it from the JTWPA gain for the offset of ON data, and by adding it to the line attenuation for the offset of OFF data. Without the JTWPA in the chain, the OFF noise floor would lie at $\SI{9.41}{\kelvin}$.

A convenient way to assess noise is to express it in terms of noise temperature. We can write the whole system noise power as \cite{braggio2022, vora2023}:
\begin{equation}
\label{eq:Pn}
    P_{\text{n}} = G_{\text{out}}k_{\text{B}}T_{\text{sys}}B + S_{\text{n}}^{\text{SA}}B
\end{equation}
with $G_{\text{out}} = +76.1 \pm \SI{1}{\decibel}$ the full output gain after the device, $S_\text{n}^{\text{SA}}=\SI{-142}{\deci\belmilliwatt\per\hertz}$ the noise power density of the SA, and $T_\text{sys}$ the system noise temperature. The noise floor JTWPA ON on bare SA measurement reads $P_{\text{n}} = \SI{-97.8}{\deci\belmilliwatt}$. From this, we calculate a system noise temperature of $T_\text{sys} = 0.98^{+0.3}_{-0.2} \SI{}{\kelvin}$. Expressed in terms of measurement quantum efficiency, we have $\eta = hf/k_\text{B}T_\text{sys} = 14\%$. This low value is explained by the limited JTWPA gain which is not enough to overcome the subsequent HEMT added noise, by the high ($\SI{4}{\kelvin}$) HEMT added noise temperature compared to state-of-the-art HEMTs, and by additional losses in RF circuitry before the HEMT.

To disentangle these different contributions, and to investigate the JTWPA part in this limited measurement efficiency, we rely on the following model for noise estimation in an amplification chain \cite{blais2021}, illustrated in Fig.\ref{fig:noise}(c). As the standard quantum limit of noise added by an amplification process is conveniently given by half a quantum of energy at the frequency of interest \cite{Caves1982}, the following discussion expresses noise in numbers of quanta. It is then useful to recall that the number of noise quanta emitted by a thermal source at frequency $f$ and temperature $T$ at equilibrium is given by the Bose-Einstein distribution:
\begin{equation}
\label{eq:thermal_noise}
    N(f,T) = \frac{1}{\exp\left( \frac{hf}{k_\text{B}T} \right) - 1}
\end{equation}
We model the RF circuit between the device and the JTWPA, and the RF circuit between the JTWPA and the HEMT, as lossy beam splitters with transmissivities $\eta_{1,2}$. In this situation, the whole system consisting of the two amplifiers and two beam splitters is equivalent to one amplifier with total gain and added noise:
\begin{align}
    \label{eq:noise_model}
    G_T &= \eta_1\eta_2G_{\text{JTWPA}}G_{\text{HEMT}} \\ \nonumber
    N_T &= \frac{1}{G_T-1}\left[ \eta_1(G_{\text{JTWPA}}- 1)G_{\text{HEMT}}(N_{\text{JTWPA}}+1) \right. \\ \nonumber
    & +\left. (G_{\text{HEMT}}-1)(N_{\text{HEMT}}+1)\right] - 1
\end{align}
We make the approximation that the gain of the HEMT is high enough to discard the added noise coming from further components of the amplification chain. Thus, the noise measured by the SA renormalized by the amplification chain up to point "REF" in Fig.\ref{fig:noise}(c) is this total added noise $N_T$, plus the incoming noise $N_{\text{ref}}$ at point "REF". $N_\text{ref}$ sums up the noise coming from the device line and from the pump line. The one coming from the pump line is attenuated by $\SI{20}{\decibel}$ at the cold plate, and further by $\SI{18}{\decibel}$ in the directional coupler, so it is considered negligible. The noise at the output of the coupler is then considered thermal noise at frequency $f$ and temperature $T=\SI{50}{\milli\kelvin}$ at equilibrium, so $N_\text{ref}=0.06$ quanta from eq.(\ref{eq:thermal_noise}). $N_\text{sys} = k_\text{B}T_\text{sys}/hf$ is computed from eq.(\ref{eq:Pn}), with $G_\text{out} = G_{T} \times G_\text{RT}$ with $G_\text{RT}$ the gain of the room temperature RF circuitry. Finally, writing $N_\text{sys} = N_T + N_\text{REF}$ and isolating the JTWPA added noise, we find $N_\text{JTWPA} = 0.89$ energy quantum, close to the $0.5$ quantum limit. The difficulty to estimate the gains and attenuations implies that we can not assert this number precisely, but we are definitely in the near-quantum limited regime. This JTWPA added noise performance can be explained by the reduced dielectric losses when going to lower frequencies.

\section{V. SNR improvement}

\begin{figure}%
\includegraphics[width=\columnwidth]{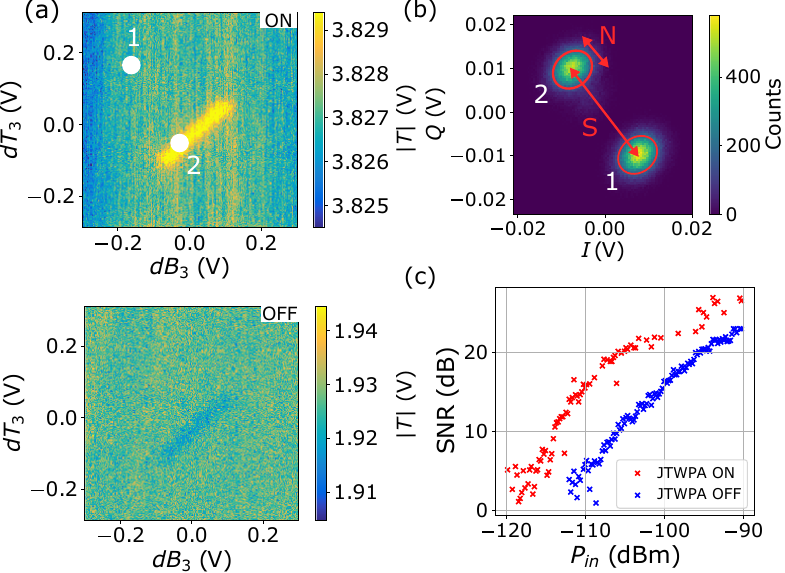}
\caption{JTWPA SNR improvement of RF dispersive measurements of interdot charge transitions. (a) Typical RF output signal at a charge transition between the two quantum dots below gates $\text{T}_3$ and $\text{B}_3$. We show here only the transmission modulus quadrature. Point 1 is a gate voltage point of charge stability. Point 2 is a gate voltage configuration enabling electron tunneling. This tunneling possibility adds up a capacitance signal in the resonator, shifting the resonance frequency and thus giving a different transmission amplitude. (b) $(I,Q)$-plot of the RF output signal. Data at the 1 and 2 positions of (a) are binned in a two-dimensional histogram. The two resulting blobs are fitted to tilted gaussians. The signal $S$ is defined as the distance between their centers, the noise $N$ is the gaussian half width at half maximum, and $\text{SNR}=S/N$. (c) SNR computed as explained in (b) on a different device, and studied as a function of input power offsetted by input line attenuation $P_\text{in}$ up to the the JTWPA input, comparing JTWPA ON (red) and OFF (blue).
}
\label{fig:snr}
\end{figure}

We can now turn to impedance measurements of interdot charge transitions with the help of the JTWPA. Fig.\ref{fig:snr}(a) shows a comparison of the RF transmitted signal $|T|$ between JTWPA ON and OFF, at a gate voltage configuration where an electron is able to tunnel between dots formed under $T_3$ and $B_3$, measured in the setup described in Fig.\ref{fig:setup}(a).

To quantify the SNR improvement provided by the JTWPA, we represent the interdot feature in the $(I,Q)$ plane in Fig.\ref{fig:snr}(b). To do so, we toggle between the two positions in the stability diagram corresponding to a fixed charge (point "1" in the figure), and to a charge degeneracy point on an interdot signal (point "2"). We bin these $(I,Q)$ data and fit them to a two-dimensional double gaussian curve (see Appendix 2). The SNR is then computed as the ratio of the distance between the gaussian centers (signal $S$), and the mean extension of these gaussian bells (noise $N$). In this representation, the effect of the JTWPA is a better separation of the gaussians, without enlarging them.

To show evidence of reproducibility and applicability of the JTWPA characteristics of the previous section on this signal of interest, we repeat the experiment described in Fig.\ref{fig:snr}(b) on a similar device consisting of only one pair of split gates. The resonance frequency of the corresponding resonator was $\SI{3.32}{\giga\hertz}$, similar to the device described above. We retuned the JTWPA pump frequency and power to maximize the gain at this readout position, giving a pump frequency of $\SI{4.921}{\giga\hertz}$ and power $\SI{-6.83}{\deci\belmilliwatt}$. Fig.\ref{fig:snr}(c) shows SNR data computed from $(I,Q)$ plots as described just before, for a power incoming into the device from $\SI{-120}{\dBm}$ to $\SI{-90}{\dBm}$. We first verify the $\text{SNR} \propto P_\text{in}$ as in relation (\ref{eq:snr}). This plot contains different zones. Below $\SI{-110}{\dBm}$, the JTWPA makes the signal stand out from the noise, when there is no consistent SNR value for JTWPA OFF data. Between $\SI{-110}{\dBm}$ and $\SI{-100}{\dBm}$ we get around $\SI{10}{\decibel}$ SNR improvement, consistent again with the two other devices studied in this work. This performance finally drops off after around $\SI{-100}{\dBm}$. This SNR improvement saturation is consistent with standard compression point measurements shown in Fig.\ref{fig:JTWPA_charac}(c) for the three split-gates device. This result illustrates that the direct measurements of the JTWPA characteristics are found unchanged when the amplifier is used in typical semiconductor quantum dot experiments. In particular, it can be noted here that up to this point, we reproduced a stable ON/OFF SNR improvement of around $+\SI{10}{\decibel}$ over three similar SiMOS devices, with resonance frequencies ranging from $\SI{2.911}{\giga\hertz}$ to $\SI{3.525}{\giga\hertz}$.

\begin{figure}%
\includegraphics[width=\columnwidth]{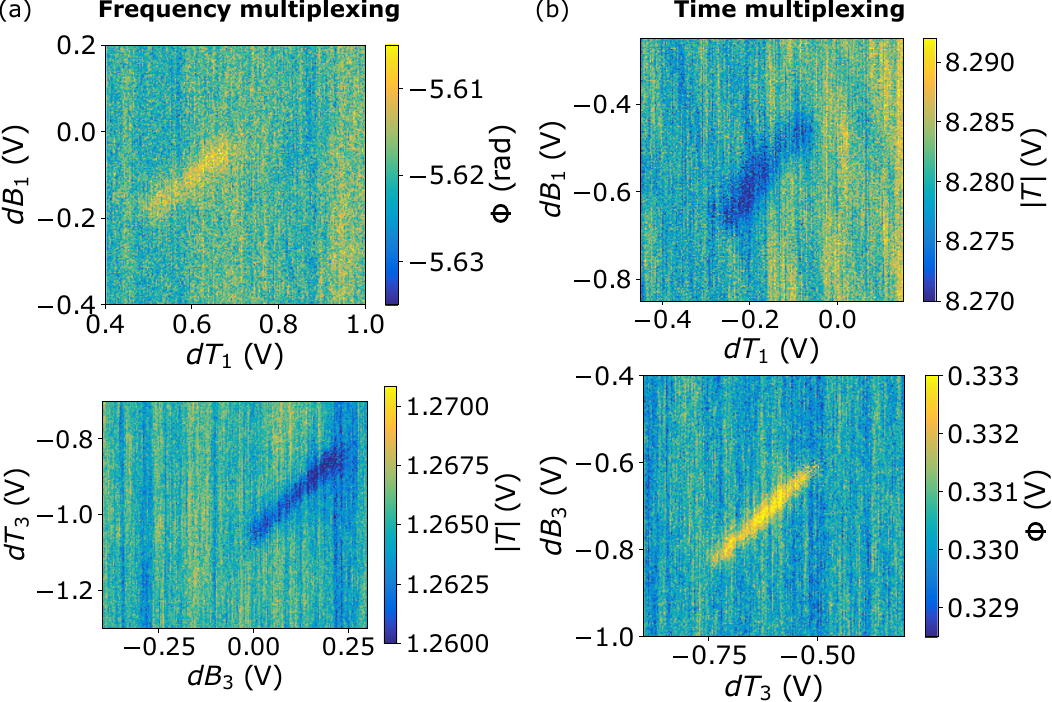}
\caption{Multiplexed readout of two interdot charge transitions. Top panels show $(\text{B}_1,\text{T}_1)$ transition, bottom panels show $(\text{B}_3,\text{T}_3)$ transition. We only show the cleanest quadrature between modulus and phase. (a) Frequency-division multiplexing data. (b) Time-division multiplexing data.
}
\label{fig:mux}
\end{figure}

\section{VI. Multiplexed readout}

Finally, we investigate multiplexed readout of two interdot charge transitions. Going back to the three split-gates device, we first implement frequency multiplexing of the readout of an interdot charge transition in the DQD below $\text{T}_1$ and $\text{B}_1$, and of an interdot transition in the DQD below $\text{T}_3$ and $\text{B}_3$. The experimental setup differs from the one described on Fig.\ref{fig:setup}(a) at room temperature: here, the RF source is used to generate a $\SI{3.46}{\giga\hertz}$ carrier frequency. A second source is used to generate two modulation signals at $f^\text{m}_1=\SI{5}{\mega\hertz}$ and $f^\text{m}_3=\SI{50}{\mega\hertz}$, which are then added together and mixed with the carrier signal thanks to a single-sideband modulator (SSB). The resulting signal is then sent into the cryostat. It is worth noting here that the tone separation is limited by the SSB $5-\SI{50}{\mega\hertz}$ IF bandwidth, forcing us to work at non-optimal frequencies for reflectometry. The output RF signal from the cryostat is first down-converted by the carrier frequency, and the resulting signal is demodulated at $f^\text{m}_1$ and $f^\text{m}_3$.

Fig.\ref{fig:mux}(a) shows both interdots measured simultaneously. $\text{B}_1$ and $\text{T}_3$ are fast-swept together along the $y$-axis, and $\text{T}_1$ and $\text{B}_3$ (the two gates connected to a RF circuit) are stepped together along the $x$-axis. The top panel shows the phase of the signal demodulated at $f^\text{m}_1$ with interdot corresponding to ($\text{T}_1/\text{B}_1$) DQD, and the bottom panel shows the modulus of the signal demodulated at $f^\text{m}_3$ with interdot corresponding to ($\text{T}_3/\text{B}_3$) DQD. We always measure both the transmission modulus and phase, but the figure shows only the quadrature containing the cleanest signal, see Appendix 3 for the complete measurement data. Beside the SSB bandwidth constraint, the quality of this frequency-multiplexed signal is mainly limited because of the quality factor of the resonators, which degrades the SNR, and additionally creates cross-talk between the two signals of interest. Also, both tones must be $\SI{3}{\decibel}$ lower than individual ones, in order to work at maximum $P_\text{in}$ but staying below JTWPA compression. 

Another alternative to perform multiple readout while maintaining the total power below compression is to perform time- rather than frequency-multiplexing. Fig.\ref{fig:mux}(b) shows interdot transitions below ($\text{T}_1/\text{B}_1$) (top panel, measured at $f_1=\SI{3.385}{\giga\hertz}$) and below ($\text{T}_3/\text{B}_3$) (bottom panel, measured at $f_3=\SI{3.565}{\giga\hertz}$). Again, we kept only the best quadratures. Both maps are taken together in an alternating fashion: for each point, we toggle between $\SI{2}{\milli\second}$ at a gate voltage configuration corresponding to a point on the $\text{T}_1/\text{B}_1$ map and input frequency $f_1$ while $\text{T}_3/\text{B}_3$ are at $\SI{0}{\volt}$, and $\SI{2}{\milli\second}$ on the $\text{T}_3/\text{B}_3$ map and input frequency $f_3$ while $\text{T}_1/\text{B}_1$ are at $\SI{0}{\volt}$, with $f_1$ and $f_3$ the optimal reflectometry frequencies for both transitions. We toggle between $f_1$ and $f_3$ thanks to a fast-gated RF switch. We retrieve both interdot features with the JTWPA ON during all the duration of experiments, with no need to retune the pump parameters. This was enabled by the broad $\SI{500}{\mega\hertz}$ band of JTWPA gain in this pump configuration, and opens the path towards multiplexing of a larger number of resonators.

\section{VII. Conclusion}
In this paper, we demonstrated the convenient and reproducible use of a single Josephson junction based superconducting traveling-wave parametric amplifier (JTWPA) in dispersive gate-based radiofrequency (RF) readout of interdot charge transitions in SiMOS quantum dots.

The JTWPA offers more than $\SI{10}{\decibel}$ signal-to-noise ratio improvement over a broad $\SI{2}{\giga\hertz}$ bandwidth, with added noise close to the quantum limit, comparable to its typical characteristics in usual working regimes for superconducting qubit experiments. We demonstrated an advantage of using the JTWPA on interdot charge transition signals in gate-defined corner dots in SiMOS structures over a $\SI{30}{\decibel}$ range of input power, and its reproducibility on three different devices. We leveraged its broad amplification band together with a high $\SI{-100}{\deci\belmilliwatt}$ $\SI{1}{\decibel}$-compression point to multiplex the readout of interdot charge transitions between two separate double-quantum dots in the same multi-gate SiMOS device, in both frequency and time domains, measured with a method equivalent to gate-based radiofrequency reflectometry around $\SI{3}{\giga\hertz}$, thanks to low footprint lumped-element resonators post-processed directly on the SiMOS chip. Additionally, we showed the JTWPA magnetic field resilience up to $\SI{1.2}{\tesla}$ felt by the quantum dot device, resulting simply from the single-junction nature of the superconducting chain, from its metallic top ground, and from its position and orientation in the cryostat. This lowers the cost and effort compared to usual mu-metal shielding necessary for SQUID-based amplifiers, and opens the road towards multiplexed spin qubit readout in SiMOS devices. These results highlight the usefulness of such JTWPAs in the context of semiconductor spin qubit experiments, especially in the context of scalable quantum computation. This comes at the cost of additional lossy circuitry to link the JTWPA to the device and to the HEMT, resulting in a reduced overall measurement quantum efficiency.

Nevertheless, the signal quality and the measurement integration time do not compare with state-of-the-art semiconductor quantum dots experiments using lambda resonators and similar superconducting amplifier performance \cite{macklin2015, dejong2021}. In order to bridge this gap, different elements need improvement. With further lumped-element resonator engineering, better quality factors would enable to reduce readout cross-talk and improve the dispersive shift part of the SNR \cite{vonhorstig2023}. Working on the device design could give a finer knowledge and control over interdot transition rates \cite{Niebojewski2022}, to which we could adapt the frequency and power of the RF readout method. In both regards, co-integration of lumped-element resonators and SiMOS devices on-chip will alleviate the need for bonding pads and wires, giving reduced resonator footprint and parasitic capacitance. Finally, on the superconducting amplifier side, adapting the photonic gap position and depth could enable to lower the attainable frequency regime and get to less uncharted bands for lumped element RF resonators, where better quality resonances are routinely achieved.

\section{Acknowledgments}

We acknowledge technical support from D. Dufeu, G. Bres, P. Carecchio, T. Crozes, D. Lepoittevin, L. Hutin, I. Pheng, L. Del Rey, J. Jarreau, C. Hoarau and C. Guttin. We thank M. Nurizzo, B. Jadot, R. Maurand and G. Le Gal for fruitful discussions, and J. Renard for his help with the cryogenic microwave setup.

The device fabrication is funded through the Mosquito project (Grant agreement No.688539) and this experimental work is funded through the QuCube project (Grant agreement No.810504) and the QLSI project (Grant agreement No.951852).

\section{Authors contributions}
H.N, B.B and M.V were responsible for the front-end fabrication of the SiMOS devices. B.K, V.T and R.L developed the post-processing of the inductor and contacts on the SiMOS device. E.E designed, fabricated and installed the cryostat. L.P developed and fabricated the JTWPA for the purpose of this study. V.E carried out the experiments with help from M.U, D.N, B.C, A.B, L.P and M.D. V.E wrote the manuscript with inputs from all the authors. M.U, N.R and T.M supervised and initiated the project. 

\section{Competing Interests}
The authors declare no competing financial or non-financial interests.

\section{Data availability}
Data that support this study are available from the corresponding authors upon request.

\newpage 
\clearpage

\section{Appendix}

\appendix

\section{1. Noise performance estimation}

In this appendix, we complete the discussion about the estimation of the noise performance.

\begin{figure}%
\includegraphics[width=\columnwidth]{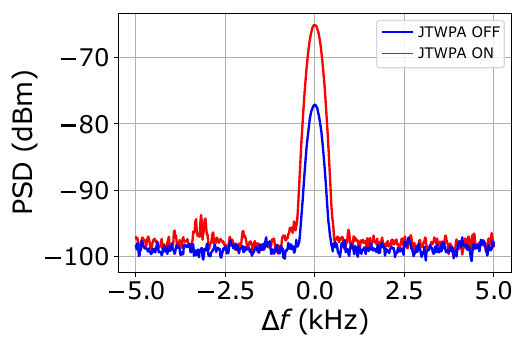}
\caption{(a) Bare spectrum analyzer measurements corresponding to main Fig.\ref{fig:JTWPA_charac}(c). (b) RF transmission with JTWPA unpumped but plugged, and with JTWPA replaced with a $\SI{50}{\ohm}$ through. We measure a $\SI{4}{\decibel}$ insertion loss at $f = \SI{2.911}{\giga\hertz}$. (c) Model schematic of the RF circuit for noise estimation. The output of the directional coupler adds up the signal coming from the device, and the signal coming from the pump attenuated by $\SI{18}{\decibel}$. The point "REF" is the reference point for noise calculations, in order to express all the noise contributions as referred to the JTWPA input. $\eta_{1,2}$ and $N_{1,2}$ are the transmissivity and vacuum noise of the lossy beam splitters that model circuitry between the coupler and the JTWPA, and between the JTWPA and the HEMT.
}
\label{fig:noise_appendix}
\end{figure}

First, Fig.\ref{fig:noise_appendix} shows bare spectrum analyzer (SA) data corresponding to main Fig.\ref{fig:noise}(a), without renormalization by the gain of the entire RF output chain. The noise power value of $P_{\text{n}} = \SI{-97.8}{\deci\belmilliwatt}$ in the main text is the mean of the JTWPA ON curve outside from the peak.

Then, a delicate point is to estimate precisely each element of the amplification chain, in order to compute the system noise temperature with eq.(\ref{eq:Pn}), and the JTWPA added noise with eq.(\ref{eq:noise_model}). As we do not have access to the RF circuitry when the cryostat is running, we estimate gains and attenuations from constructor specifications, see table \ref{tab:output_estim}. We realize these numbers can differ from real performance, and this is the reason why we do not claim a precise $N_\text{JTWPA} = 0.89$ added noise photon number. But the reduced dielectric loss due to lower operating frequencies should result in a better noise performance than in nominal operating regimes, which is consistent with our estimation. We also note that the discrepancy between this apparently very good JTWPA noise performance and our limited 14\% measurement quantum efficiency can be explained by the $\SI{-2.58}{\decibel}$ losses in RF circuitry between the JTWPA and the HEMT amplifier, and by the high added noise of the HEMT.

\begin{table}[t]
\caption{Estimation of the gain or attenuation of the different RF components in the output chain.}
\label{tab:output_estim}
\begin{tabular}{cc}
\hline
RF component & Gain / Attenuation (dB) \\ \hline
SMA before JTWPA & -0.37 \\
SMA after JTWPA & -0.37 \\
Isolator & -0.2 \\
SMA after isolator & -0.13 \\
SMA from $\SI{50}{\milli\kelvin}$ to HEMT & -1.88 \\
HEMT & +36 \\
$\SI{300}{\kelvin}$ amplifiers & +32.2\\
\hline
\end{tabular}
\end{table}

\section{2. IQ plane SNR computation}

In this section, we explain the way we compute signal-to-noise ratio (SNR) in the readout of interdot charge transitions (main Fig.\ref{fig:snr}). In the $(I,Q)$ plane, we bin measurement points taken in a region of charge stability (background of a stability diagram), and measurement points taken on an interdot charge transition, where the charge degeneracy adds a capacitance contribution to the resonance frequency and thus shifts the transmitted signal $I$ and $Q$ output values. We obtain a histogram of two blobs, to which we can fit the following double gaussian shaped curve:

\begin{align}
    f(i,q) = & \nonumber \\
    A_1\times \exp [ -& \left( \cos(\theta_1)(i_1-i) + \sin(\theta_1)(q_1-q) \right)^2 / (2\sigma_{1x})^2 \nonumber \\
                            -& \left( (-\sin(\theta_1))(i_1-i) + \cos(\theta_1)(q_1-q) \right)^2 / (2\sigma_{1y})^2 ] \nonumber \\
    + A_2\times \exp [ -& \left( \cos(\theta_2)(i_2-i) + \sin(\theta_2)(q_2-q) \right)^2 / (2\sigma_{2x})^2 \nonumber \\
                            -& \left( (-\sin(\theta_2))(i_2-i) + \cos(\theta_2)(q_2-q) \right)^2 / (2\sigma_{2y})^2 ]
\end{align}
Fit parameters are:
\begin{itemize}
    \item $(i_1,q_1)$ and $(i_2,q_2)$ the positions of the centers of the blobs \\
    \item $\theta_1$ and $\theta_2$ the tilt of the blobs compared to $I$ and $Q$ axes \\
    \item $A_1$ and $A_2$ the amplitudes of the gaussians
    \item $(\sigma_{1x},\sigma_{1y})$ and $(\sigma_{2x},\sigma_{2y})$ the 2-dimensional extensions of these gaussians along the blobs axes
\end{itemize}
The SNR quantifies the ease to tell apart the two blobs. The signal is then computed as the distance between the gaussian centers, and the noise is their mean extension (we take the highest of both):
\begin{align}
    S &= \sqrt{(q_1-q_2)^2 + (i_1-i_2)^2} \\
    N &= \max\left(\frac{\sigma_{1x} + \sigma_{1y}}{2}, \frac{\sigma_{2x} + \sigma_{2y}}{2}\right) \\
    \text{SNR} &= S/N
\end{align}

\section{3. Frequency multiplexing complete data}

\begin{figure}%
\includegraphics[width=\columnwidth]{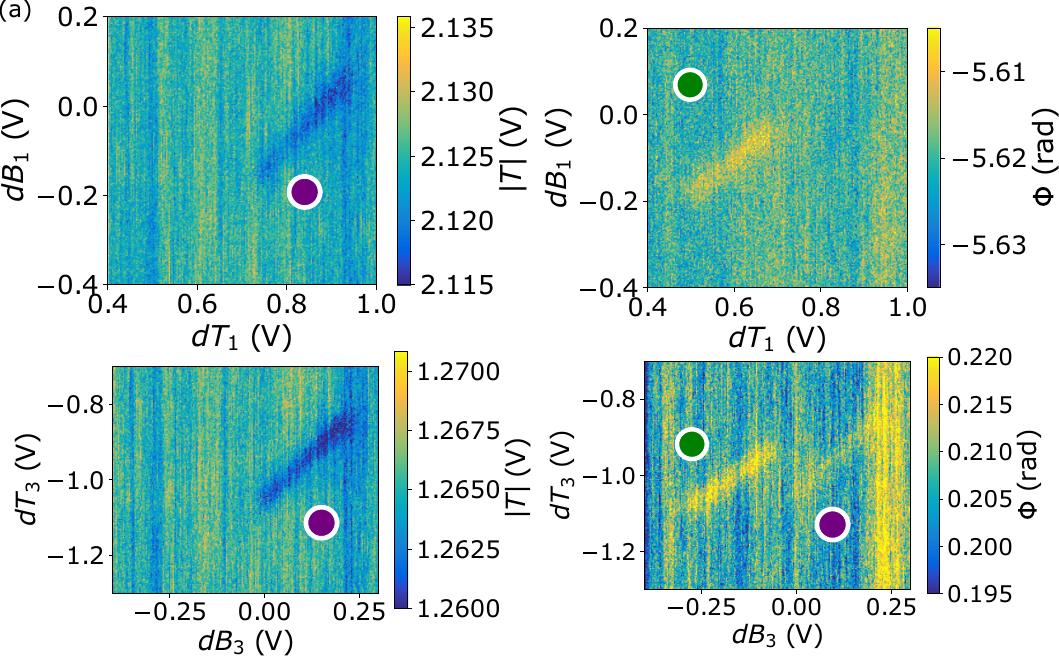}
\caption{Complete frequency multiplexing data. Top panels show the signals demodulated at $f^\text{m}_1$ in the main text, and bottom panels at $f^\text{m}_3$. Left panels show the transmission modulus, and right panels the phase. We pinpoint with green spots the interdot charge transition corresponding to ($\text{T}_1/\text{B}_1$) DQD, and with purple spots the interdot charge transition corresponding to ($\text{T}_3/\text{B}_3$) DQD.
}
\label{fig:freqmux_appendix}
\end{figure}

In main Fig.\ref{fig:mux}, we chose to show only the signal quadrature giving the best result for frequency multiplexing purposes. Here in Fig.\ref{fig:freqmux_appendix}, we show both quadrature (transmission modulus $|T|$ in left panels and phase $\Phi$ in right panels) of signals demodulated at both modulation frequencies ($f^\text{m}_1$ in upper panels and $f^\text{m}_3$ in lower panels). We pinpoint with green spots the interdot charge transition corresponding to the ($\text{T}_1/\text{B}_1$) DQD, and with purple spots the transition of ($\text{T}_3/\text{B}_3$) DQD.

We first get expected frequency multiplexing results: the  phase of the signal demodulated at $f^\text{m}_1$ (upper right panel) contains the ($\text{T}_1/\text{B}_1$) interdot, and the modulus of the signal demodulated at $f^\text{m}_3$ (lower left panel) contains the ($\text{T}_3/\text{B}_3$) interdot.

But we also get unwanted signals from these measurements. The modulus of the signal demodulated at $f^\text{m}_1$ (upper left panel) contains the ($\text{T}_3/\text{B}_3$) interdot, while it should show the other interdot at this demodulation frequency. And the phase of the map demodulated at $f^\text{m}_3$ (lower right panel) even shows both interdots at the same time. These parasitic signals most likely comes from leakage between resonators, given their poor quality factors while having similar resonance frequencies.

\end{document}